\documentclass[conference]{IEEEtran}
\usepackage{graphicx}
\usepackage{amsmath}
\usepackage{array}
\newcommand{\PreserveBackslash}[1]{\let\temp=\\#1\let\\=\temp}
\newcolumntype{C}[1]{>{\PreserveBackslash\centering}p{#1}}
\newcolumntype{R}[1]{>{\PreserveBackslash\raggedleft}p{#1}}
\newcolumntype{L}[1]{>{\PreserveBackslash\raggedright}p{#1}}
\usepackage[usenames]{color}
\usepackage{colortbl,booktabs}
\usepackage{tabularx}
\usepackage{multicol}
\usepackage{booktabs}
\usepackage[Symbol]{upgreek}
\usepackage{subfigure}
\usepackage{bm}
\usepackage{cite}
\usepackage{balance}
\usepackage[boxed]{algorithm2e}

\usepackage{threeparttable}
\usepackage{amsmath}
\usepackage{dcolumn}
\usepackage{multirow}
\usepackage{booktabs}
\usepackage{amsfonts}
\newcolumntype{d}[1]{D{.}{.}{#1}}

\def \qed {\hfill \vrule height6pt width 6pt depth 0pt}

\begin{document}

\bibliographystyle{IEEEtran} 
\title{Machine Learning Inspired Energy-Efficient Hybrid Precoding for MmWave Massive MIMO Systems}

\author{
\IEEEauthorblockN{Xinyu Gao$^\ast$, Linglong Dai$^\ast$, Ying Sun$^\ast$, Shuangfeng Han$^\dagger$, and Chih-Lin I$^\dagger$}
\IEEEauthorblockA{$^\ast$Tsinghua National Laboratory for Information Science and Technology (TNList), \\Department of Electronic Engineering, Tsinghua University, Beijing, China\\
$^\dagger$Green Communication Research Center, China Mobile Research
Institute, Beijing 100053, China}
}

\maketitle
\begin{abstract}
Hybrid precoding is a promising technique for mmWave massive MIMO systems, as it can considerably reduce the number of required radio-frequency (RF) chains without obvious performance loss. However, most of the existing hybrid precoding schemes require a complicated phase shifter network, which still involves high energy consumption. In this paper, we propose an energy-efficient hybrid precoding architecture, where the analog part is realized by a small number of switches and inverters instead of a large number of high-resolution phase shifters. Our analysis proves that the performance gap between the proposed hybrid precoding architecture and the traditional one is small and keeps constant when the number of antennas goes to infinity. Then, inspired by the cross-entropy (CE) optimization developed in machine learning, we propose an adaptive CE (ACE)-based hybrid precoding scheme for this new architecture. It aims to adaptively update the probability distributions of the elements in hybrid precoder by minimizing the CE, which can generate a solution close to the optimal one with a sufficiently high probability. Simulation results verify that our scheme can achieve the near-optimal sum-rate performance and much higher energy efficiency than traditional schemes.
\end{abstract}

\section{Introduction}\label{S1}
\IEEEPARstart Millimeter-wave (mmWave) massive multiple-input multiple-output (MIMO) has been considered as a promising technology for future 5G wireless communications~\cite{swindlehurst2014millimeter}, since it can provide wider bandwidth~\cite{rappaport2013millimeter} and achieve higher spectral efficiency~\cite{marzetta10}. However, in MIMO systems, each antenna usually requires a dedicated radio-frequency (RF) chain (including high-resolution digital-to-analog converter, mixer, etc.) to realize the fully digital signal processing (e.g., precoding)~\cite{xie2016unified}. For mmWave massive MIMO, this will result in unaffordable hardware complexity and energy consumption, as the number of antennas becomes huge and the energy consumption of RF chain is high~\cite{heath2015overview}. To reduce the number of required RF chains, hybrid precoding has been recently proposed~\cite{el2013spatially}. Its key idea is to decompose the fully digital precoder into a large-size analog beamformer (realized by the analog circuit) and a small-size digital precoder (requiring a small number of RF chains). Thanks to the low-rank characteristics of mmWave channels~\cite{rappaport2013millimeter}, a small-size digital precoder can achieve the spatial multiplexing gains, making hybrid precoding enjoy the near-optimal performance~\cite{heath2015overview}.

Nevertheless, most of the existing hybrid precoding schemes require a complicated phase shifter network, where each RF chain is connected to all antennas with high-resolution phase shifters~\cite{el2013spatially,xie2016overview}. Although this architecture can provide high design freedom to achieve the near-optimal performance, it requires hundreds or even thousands of high-resolution phase shifters with high hardware cost and energy consumption~\cite{heath2015overview}. To solve this problem, two categories of schemes have been proposed very recently. The first category is to directly employ finite-resolution phase shifters instead of high-resolution phase shifters~\cite{sohrabi2015hybrid,alkhateeb2015limited}. It can reduce the energy consumption of phase shifter network without obvious performance loss, but it still requires a large number of phase shifters with considerable energy consumption. The second category is to utilize the switch network to replace the phase shifter network~\cite{alkhateeb2016massive,mendez2015channel,sayeed2013beamspace}. It can significantly reduce the hardware cost and energy consumption, but it suffers from an obvious performance loss.

In this paper, we propose a switch and inverter (SI)-based hybrid precoding architecture with considerably reduced hardware cost and energy consumption. Instead of using phase shifters, the analog part of the proposed architecture is realized by a small number of energy-efficient switches and inverters. Then, we provide the performance analysis to quantify the performance gap between the proposed hybrid precoding architecture and the traditional ones. After that, inspired by the cross-entropy (CE) optimization developed in machine learning~\cite{rubinstein2013cross}, we  propose an adaptive CE (ACE)-based hybrid precoding scheme for this new architecture. Specifically, according to the probability distributions of the elements in hybrid precoder, this scheme first randomly generates several candidate hybrid precoders. Then, it adaptively weights these candidate hybrid precoders according to their achievable sum-rates, and refines the probability distributions of elements in hybrid precoder by minimizing the CE. Repeating such procedure, we can finally generate a hybrid precoder close to the optimal one with a sufficiently high probability. Simulation results verify that our scheme can achieve the near-optimal sum-rate performance and much higher energy efficiency than traditional schemes.


{\it Notation}: Lower- and upper-case boldface letters denote vectors and matrices, respectively;  ${( \cdot )^T}$, ${( \cdot )^H}$, ${( \cdot )^{ - 1}}$, ${{\rm{tr}}\left(  \cdot  \right)}$, and ${{\left\|  \cdot  \right\|_F}}$ denote the transpose, conjugate transpose, inversion, trace, and Frobenius norm of a matrix, respectively; ${\left|  \cdot  \right|}$ denotes the absolute operator; ${\mathbb{E}( \cdot )}$  denotes the expectation; ${ \otimes }$ denotes the kronecker product; ${{\bf{I}}_N}$ is the  $ N \times N $  identity matrix.

\section{System Model}\label{S2}
In this paper, we consider a typical mmWave massive MIMO system, where the base station (BS) employs ${N}$ antennas and ${{N_{{\rm{RF}}}}}$ RF chains to simultaneously serve ${K}$  single-antenna users (the extension to users with multiple-antennas is also possible as will be explained later). To fully achieve the multiplexing gains, we assume ${{N_{{\rm{RF}}}} = K}$~\cite{alkhateeb2015limited}. For hybrid precoding systems as shown in Fig. 1, the  ${K \times 1}$ received signal vector ${{\bf{y}}}$ for all ${K}$ users can be presented by
\begin{equation}\label{eq1}
{\bf{y}} = {\bf{H}}{{\bf{F}}_{{\rm{RF}}}}{{\bf{F}}_{{\rm{BB}}}}{\bf{s}} + {\bf{n}},
\end{equation}
where ${{\bf{H}} = {\left[ {{{\bf{h}}_1},{{\bf{h}}_2}, \cdots ,{{\bf{h}}_K}} \right]^H}}$ is the channel matrix with ${{{\mathbf{h}}_{k}}}$ presenting the ${N \times 1}$ channel vector between the BS and the ${k}$th user, ${{\bf{s}}}$ is the ${K \times 1}$ transmitted signal vector for all ${K}$  users satisfying ${\mathbb{E}\left( \mathbf{s}{{\mathbf{s}}^{H}} \right)={{\mathbf{I}}_{K}}}$, ${{{\bf{F}}_{{\rm{RF}}}}}$ of size ${N \times {N_{{\rm{RF}}}}}$ is the analog beamformer realized by analog circuit (different architectures incur different hardware constraints as will be discussed later), ${{{\bf{F}}_{{\rm{BB}}}}}$ is the baseband digital precoder of size ${{N_{{\rm{RF}}}} \times K}$ satisfying the total transmit power constraint as ${\left\| {{{\bf{F}}_{{\rm{RF}}}}{{\bf{F}}_{{\rm{BB}}}}} \right\|_F^2 = \rho}$, where ${\rho }$ is total transmit power. Finally, ${{\bf{n}}\sim{\cal C}{\cal N}\left( {0,{\sigma ^2}{{\bf{I}}_K}} \right)}$ of size ${K \times 1}$ is the additive white Gaussian noise (AWGN) vector, where ${{{\sigma ^2}}}$ presents the noise power.

For the channel vector ${{{\mathbf{h}}_{k}}}$ of the ${k}$th user, we adopt the geometric channel model to capture the characteristics of mmWave massive MIMO channels as~\cite{heath2015overview}
\begin{equation}\label{eq2}
{{\bf{h}}_k} = \sqrt {\frac{N}{L_k}} \sum\limits_{l = 1}^{{L_k}} {\alpha _k^{\left( l \right)}{\bf{a}}\left( {\varphi _k^{\left( l \right)},\theta _k^{\left( l \right)}} \right)} ,
\end{equation}
where ${{{L_k}}}$ denotes the number of paths for user ${k}$, ${{\alpha _k^{\left( l \right)}}}$ and ${{\varphi _k^{\left( l \right)}}}$ (${{\theta _k^{\left( l \right)}}}$) for ${1 \le l \le {L_k}}$ are the complex gain and azimuth (elevation) angle of departure (AoD) of the path ${l}$ for user ${k}$, ${{\bf{a}}\left( {\varphi ,\theta } \right)}$ presents the ${N \times 1}$ array steering vector. For the typical uniform planar array (UPA) with ${N_1}$ elements in horizon and ${N_2}$ elements in vertical (${N = {N_1}{N_2}}$), we have~\cite{el2013spatially}
\begin{equation}\label{eq3}
{\bf{a}}\left( {\varphi ,\theta } \right) = {{\bf{a}}_{{\rm{az}}}}\left( \varphi  \right) \otimes {{\bf{a}}_{{\rm{el}}}}\left( \theta  \right),
\end{equation}
where ${{{\bf{a}}_{{\rm{az}}}}\left( \varphi  \right) = \frac{1}{{\sqrt {{N_1}} }}{\left[ {{e^{j2\pi i\left( {{d_1}/\lambda } \right)\sin \varphi }}} \right]^T}}$ for ${i \in {\cal I}\left( {{N_1}} \right)}$, ${{{\bf{a}}_{{\rm{el}}}}\left( \theta  \right) = \frac{1}{{\sqrt {{N_2}} }}{\left[ {{e^{j2\pi j\left( {{d_2}/\lambda } \right)\sin \theta }}} \right]^T}}$ for ${j \in {\cal I}\left( {{N_2}} \right)}$, ${{\cal I}\left( n \right) = \left\{ {0,1, \cdots ,n - 1} \right\}}$, ${\lambda }$ is the signal wavelength, and ${{{d_1}}}$ (${{{d_2}}}$) is the horizontal (vertical) antenna spacing. At mmWave frequencies, we usually have ${{d_1} = {d_2} = \lambda /2}$~\cite{han2015large}.

\begin{figure}[tp]
\begin{center}
\vspace*{-4mm}\includegraphics[width=0.65\linewidth]{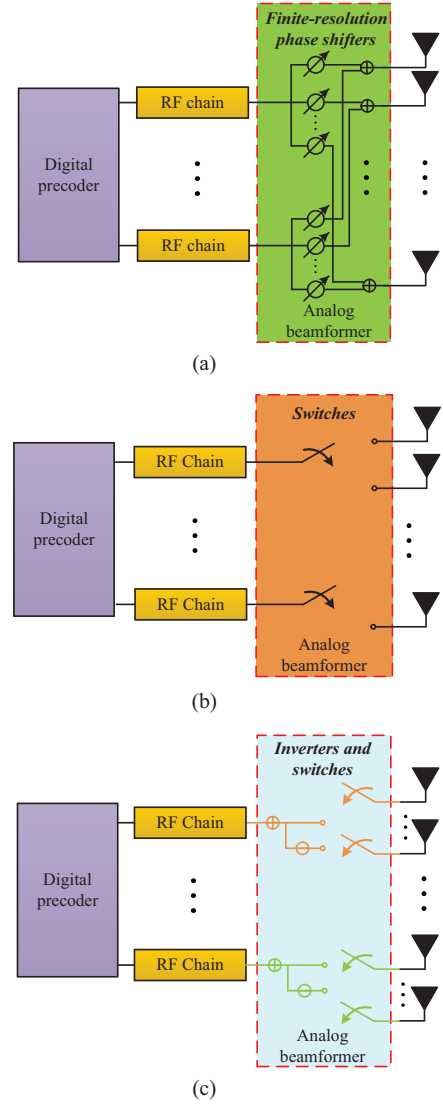}
\end{center}
\vspace*{-4mm}\caption{Hybrid precoding: (a) traditional PS-based architecture; (b) traditional SW-based architecture; (c) proposed SI-based architecture. } \label{FIG1}
\end{figure}

\section{Energy Efficient Hybrid Precoding}\label{S3}
In this section, we first describe the proposed SI-based  hybrid precoding architecture. Then, we propose an ACE-based hybrid precoding scheme for this new architecture. Finally, the complexity analysis is provided.

\subsection{The proposed SI-based hybrid precoding architecture}\label{S3.1}
Fig. 1 (a) and (b) show the traditional precoding architectures, i.e., the one with finite-resolution phase shifters (PS-based architecture)~\cite{alkhateeb2015limited} and the one with switches (SW-based architecture)~\cite{mendez2015channel}, respectively.

As shown in Fig. 1 (a), the traditional PS-based architecture requires a complicated phase shifter network, and the corresponding energy consumption can be presented as
\begin{equation}\label{eq4}
{P_{{\rm{PS-architecture}}}} = \rho  + {N_{{\rm{RF}}}}{P_{{\rm{RF}}}} + N{N_{{\rm{RF}}}}{P_{{\rm{PS}}}} + {P_{{\rm{BB}}}},
\end{equation}
where ${{P_{{\rm{RF}}}}}$, ${{P_{{\rm{PS}}}}}$, and  ${P_{{\rm{BB}}}}$ are the energy consumption of RF chain, finite-resolution phase shifter, and baseband, respectively. Note that although the PS-based architecture enjoys high design freedom to achieve the near-optimal performance~\cite{alkhateeb2015limited}, it requires a large number
(e.g., ${N \times {N_{{\rm{RF}}}} = 64 \times 16 = 1024}$~\cite{alkhateeb2015limited}) of phase shifters. Moreover, the energy consumption of finite phase shifter is also considerable (e.g., ${{P_{{\rm{PS}}}} = 40{\rm{mW}}}$ for 4-bit phase shifter~\cite{mendez2015channel}), These make the traditional PS-based architecture still suffer from high energy consumption~\cite{han2015large}.

By contrast, as shown in Fig. 1 (b), SW-based architecture can efficiently relieve such problem by employing a small number (${{N_{{\rm{RF}}}}}$ instead of ${N \times {N_{{\rm{RF}}}}}$) of energy-efficient switches. The energy consumption of SW-based architecture can be presented as
\begin{equation}\label{eq5}
{P_{{\rm{SW-architecture}}}} = \rho  + {N_{{\rm{RF}}}}{P_{{\rm{RF}}}} + {N_{{\rm{RF}}}}{P_{{\rm{SW}}}} + {P_{{\rm{BB}}}},
\end{equation}
where ${{P_{{\rm{SW}}}}}$ is the energy consumption of switch, which is much lower than ${{P_{{\rm{PS}}}}}$ (e.g., ${{P_{{\rm{SW}}}} = 5{\rm{mW}}}$~\cite{mendez2015channel}). Nevertheless, since only ${{N_{{\rm{RF}}}}}$ antennas are active simultaneously, SW-based architecture cannot fully achieve the array gains of mmWave massive MIMO, leading to an obvious performance loss~\cite{mendez2016hybrid}.

To overcome the problems faced by traditional architectures, we propose the SI-based architecture as shown in Fig. 1 (c), which can be considered as a better trade-off between the near-optimal PS-based architecture and the energy-efficient SW-based architecture. Specifically, in the proposed SI-based architecture, each RF chain is only connected to a sub antenna array with ${M = N/{N_{{\rm{RF}}}}}$ (assumed to be an integer) antennas instead of all ${N}$ antennas~\cite{gao15energy}. Moreover, each RF chain is connected to the sub antenna array via only one inverter and ${M}$ switches instead of ${N}$ phase shifters. The energy consumption of SI-based architecture can be presented by
\begin{equation}\label{eq6}
{P_{{\rm{SI-architecture}}}}\! =\! \rho \! + \!{N_{{\rm{RF}}}}{P_{{\rm{RF}}}} \!+\! {N_{{\rm{RF}}}}{P_{{\rm{IN}}}}\! +\! N{P_{{\rm{SW}}}}\! +\! {P_{{\rm{BB}}}},
\end{equation}
where ${{P_{{\rm{IN}}}}}$ is the energy consumption of inverter. It worth pointing out that the inverters can be realized by the digital chip with the energy consumption similar to switches (i.e., ${{P_{{\rm{IN}}}} \approx {P_{{\rm{SW}}}}}$)~\cite{mendez2016hybrid}. As a result, by comparing~(\ref{eq4})-(\ref{eq6}), we can conclude that the energy consumption of the proposed SI-based architecture is much lower than that of PS-based one. Furthermore, as all antennas are used, SI-based architecture can also achieve the potential array gains of mmWave massive MIMO, which will be further proved as follows.

To do this, we need to first explain the hardware constraints induced by the proposed SI-based architecture, which are different from those of the traditional ones. The first constraint is that the analog beamformer ${{{{\bf{F}}_{{\rm{RF}}}}}}$ should be a block diagonal matrix instead of a full matrix as
\begin{equation}\label{eq7}
{{\bf{F}}_{{\rm{RF}}}} = {\left[ {\begin{array}{*{20}{c}}
{{\bf{\bar f}}_1^{{\rm{RF}}}}&{\bf{0}}& \ldots &{\bf{0}}\\
{\bf{0}}&{{\bf{\bar f}}_2^{{\rm{RF}}}}&{}&{\bf{0}}\\
 \vdots &{}& \ddots & \vdots \\
{\bf{0}}&{\bf{0}}& \ldots &{{\bf{\bar f}}_{{N_{{\rm{RF}}}}}^{{\rm{RF}}}}
\end{array}} \right]_{N \times {N_{{\rm{RF}}}}}},
\end{equation}
where ${{{\bf{\bar f}}_n^{{\rm{RF}}}}}$ is the ${M \times 1}$ analog beamformer on the ${n}$th sub antenna array. The second constraint is that since only inverters and switches are used, all the ${N}$ nonzero elements of ${{{{\bf{F}}_{{\rm{RF}}}}}}$ should belong to
\begin{equation}\label{eq32}
\frac{1}{{\sqrt N }}\left\{ { - 1, + 1} \right\}.
\end{equation}
Based on these constraints, we have the following \textbf{Lemma 1}.

\vspace*{+2mm} \noindent\textbf{Lemma 1}. {\it Assume that the channel ${{{\bf{h}}_k}}$ of user ${k}$ only has single path, i.e., ${{L_k} = 1}$~\cite{sayeed2013beamspace}. When ${N \to \infty }$ and ${N/M = {N_{{\rm{RF}}}}}$, the ratio ${\zeta }$ between the array gains achieved by SI-based architecture and that achieved by PS-based architecture with sufficiently high-resolution phase shifters can be presented by}
\begin{equation}\label{eq21}
\mathop {\lim }\limits_{N \to \infty, \frac{N}{M} = {N_{{\rm{RF}}}}} \zeta  = \frac{4}{{{N_{{\rm{RF}}}}{\pi ^2}}}.
\end{equation}

\textit{Proof:} For the traditional PS-based architecture with sufficiently high-resolution phase shifters, the phases of the elements in the analog beamformer can be arbitrarily adjusted to capture the power of ${{{\bf{h}}_k}}$. Therefore, the array gains achieved by PS-based architecture is ${{\left| {\alpha _k^{\left( 1 \right)}} \right|^2}}$. By contrast, the array gains achieved by SI-based architecture can be presented by
\begin{align}\label{eq22}
& {\left| {{\bf{h}}_k^H{\bf{f}}_k^{{\rm{RF}}}} \right|^2} = N{\left| {\alpha _k^{\left( 1 \right)}} \right|^2}{\left| {{{\bf{a}}^H}\left( {{\varphi _k}} \right){\bf{f}}_k^{{\rm{RF}}}} \right|^2}\\ \nonumber
& = \frac{1}{N}{\left| {\alpha _k^{\left( 1 \right)}} \right|^2}{\left| {\sum\limits_{m = 1}^M {{e^{j{{\bar \phi }_m}}}} } \right|^2}\\ \nonumber
& = \frac{1}{N}{\left| {\alpha _k^{\left( 1 \right)}} \right|^2}{\left( {{{\left| {\sum\limits_{m = 1}^M {\cos \left( {{{\bar \phi }_m}} \right)} } \right|}^2} + \left| {\sum\limits_{m = 1}^M {\sin \left( {{{\bar \phi }_m}} \right)} } \right|}^2 \right)},
\end{align}
where ${{{\bf{f}}_k^{{\rm{RF}}}}}$ is the ${k}$th column of ${{{\bf{F}}_{{\rm{RF}}}}}$ including the zeros and ${{{{\bar \phi }_m}}}$ denotes the phase quantization error. Since the nonzero elements in ${{{\bf{f}}_k^{{\rm{RF}}}}}$ belong to ${\frac{1}{{\sqrt N }}\left\{ { - 1, + 1} \right\}}$, ${{\bar \phi _m}}$ can be assumed to follow the uniform distribution ${{\cal U}\left( { - \pi /2,\pi /2} \right)}$ for ${1 \le m \le M}$~\cite{alkhateeb2016massive}. Then, we have
\begin{align}\label{eq23}
\nonumber \mathop {\mathop {\lim }\limits_{N \to \infty } }\limits_{N/M = {N_{{\rm{RF}}}}} \zeta &  \!=\! \frac{1}{{{N_{{\rm{RF}}}}M}}\left\{ {{{\left| {\sum\limits_{m = 1}^M {\cos \left( {{{\bar \phi }_m}} \right)} } \right|}^2} \!+\! {{\left| {\sum\limits_{m = 1}^M {\sin \left( {{{\bar \phi }_m}} \right)} } \right|}^2}} \right\}\\ \nonumber
&= \frac{1}{{{N_{{\rm{RF}}}}}}{\left(\mathbb{E}{\left[ {\cos \left( {{{\bar \phi }_m}} \right)} \right]} \right)^2} + {\left(\mathbb{E}{\left[ {\sin \left( {{{\bar \phi }_m}} \right)} \right]} \right)^2}\\
&= \frac{4}{{{N_{{\rm{RF}}}}{\pi ^2}}},
\end{align}
which verifies the conclusion in \textbf{Lemma 1}. \qed

\textbf{Lemma 1} indicates that although the proposed SI-based architecture suffers from some loss of array gains compared to the near-optimal PS-based architecture, the performance loss keeps constant and limited, which does not grow as the number of BS antennas goes to infinity. Recalling the low energy consumption of SI-based architecture, we can conclude that our scheme is a better trade-off between the traditional architectures, which will be also verified by simulation. Next, we will design a near-optimal hybrid precoding scheme for SI-based architecture with quite different hardware constraints.

\subsection{ACE-based hybrid precoding scheme}\label{S3.2}
We aim to design the analog beamformer ${{{{\bf{F}}_{{\rm{RF}}}}}}$ and the digital precoder ${{{{\bf{F}}_{{\rm{BB}}}}}}$ to maximize the achievable sum-rate ${R}$, which can be presented as
\begin{equation}\label{eq8}
\begin{array}{l}
\left( {{\bf{F}}_{{\rm{RF}}}^{{\rm{opt}}},{\bf{F}}_{{\rm{BB}}}^{{\rm{opt}}}} \right) = \mathop {\arg \max }\limits_{{{\bf{F}}_{{\rm{RF}}}},{{\bf{F}}_{{\rm{BB}}}}} R,\\
\quad \quad \quad \quad {\rm{s}}{\rm{.t}}{\rm{.}}\quad {{\bf{F}}_{{\rm{RF}}}} \in {\cal F}{\rm{,}}\\
\quad \quad \quad \quad \quad \quad \left\| {{{\bf{F}}_{{\rm{RF}}}}{{\bf{F}}_{{\rm{BB}}}}} \right\|_F^2 = \rho,
\end{array}
\end{equation}
where ${{\cal F}}$ denotes the set with all possible analog beamformers satisfying the two constraints~(\ref{eq7}) and~(\ref{eq32}) described above, and the achievable sum-rate ${R}$ can be presented by
\begin{equation}\label{eq9}
R = \sum\limits_{k = 1}^K {{{\log }_2}\left( {1 + {\gamma _k}} \right)},
\end{equation}
where ${{{\gamma _k}}}$ presents the signal-to-interference-plus-noise ratio (SINR) of the ${k}$th user as
\begin{equation}\label{eq10}
{\gamma _k} = \frac{{{{\left| {{\bf{h}}_k^H{{\bf{F}}_{{\rm{RF}}}}{\bf{f}}_k^{{\rm{BB}}}} \right|}^2}}}{{\sum\limits_{k' \ne k}^K {{{\left| {{\bf{h}}_k^H{{\bf{F}}_{{\rm{RF}}}}{\bf{f}}_{k'}^{{\rm{BB}}}} \right|}^2} + {\sigma ^2}} }},
\end{equation}
where ${{{\bf{f}}_k^{{\rm{BB}}}}}$ is the ${k}$th column of ${{{{\bf{F}}_{{\rm{BB}}}}}}$.

It is worth pointing out that the constraints~(\ref{eq7}) and~(\ref{eq32}) on the analog beamformer ${{{{\bf{F}}_{{\rm{RF}}}}}}$ are non-convex. This makes~(\ref{eq8}) very difficult to be solved. Fortunately,  as all the ${N}$ nonzero elements of ${{{{\bf{F}}_{{\rm{RF}}}}}}$ belong to the set ${\frac{1}{{\sqrt N }}\left\{ { - 1, + 1} \right\}}$, the number of possible ${{{{\bf{F}}_{{\rm{RF}}}}}}$ is finite. Therefore,~(\ref{eq8}) can be regarded as a non-coherent combining problem~\cite{rubinstein2013cross}. To solve it, we can first select a candidate ${{{{\bf{F}}_{{\rm{RF}}}}}}$, and compute the optimal ${{{{\bf{F}}_{{\rm{BB}}}}}}$ according to the effective channel matrix ${{\bf{H}}{{\bf{F}}_{{\rm{RF}}}}}$ without non-convex constraints. After all possible ${{{{\bf{F}}_{{\rm{RF}}}}}}$'s have been searched, we can obtain the optimal analog beamformer ${{{\bf{F}}_{{\rm{RF}}}^{{\rm{opt}}}}}$ and digital precoder ${{{\bf{F}}_{{\rm{BB}}}^{{\rm{opt}}}}}$. However, such exhaustive search scheme requires to search ${{2^N}}$ possible ${{{{\bf{F}}_{{\rm{RF}}}}}}$'s and ${{{{\bf{F}}_{{\rm{BB}}}}}}$'s, which involves unaffordable complexity as ${N}$ is usually large in mmWave massive MIMO systems (e.g., ${N = 64}$, ${{2^{64}} \approx 1.84 \times {10^{19}}}$). To solve this problem, we propose an ACE algorithm, which can be considered as an improved version of the CE algorithm developed from machine learning~\cite{rubinstein2013cross}.

At first, we would like to briefly introduce the conventional CE algorithm, which is a probabilistic model-based algorithm to solve the combining problem by an iterative procedure. In each iteration, the CE algorithm first generates ${S}$ candidates (e.g., possible hybrid precoders in our problem) according to a probability distribution. Then, it computes the objective value (e.g., achievable sum-rate in our problem) of each candidate, and selects ${{{S_{{\rm{elite}}}}}}$ best candidates as ``elite"~\cite{rubinstein2013cross}. Finally, based on the selected elites, the probability distribution is updated by minimizing the CE. Repeating such procedure, the probability distribution will be refined to generate a solution close to the optimal one with a sufficiently high probability. However, although the CE algorithm has been widely used in machine learning~\cite{rubinstein2013cross}, it still has some disadvantages. One is that the contributions of all elites are treated as the same. Intuitively, the elite with better objective value should be more important when we update the probability distribution. Therefore, if we can adaptively weight the elites according to their objective values, better performance can be expected. Following this idea, we propose an ACE algorithm to solve~(\ref{eq8}).

The pseudo-code of the proposed ACE-based hybrid precoding scheme\footnote{Note that the convergence of the proposed ACE-based hybrid precoding scheme can be proved by extending the Theorem 1 in~\cite{chenefficient}.} is summarized in \textbf{Algorithm 1}, which can be explained as follows. At the beginning, we formulate the nonzero elements in ${{{\mathbf{F}}_{\text{RF}}}}$ as ${N \times 1}$ vector ${{\bf{f}} = {\left[ {{{\left( {{\bf{\bar f}}_1^{{\rm{RF}}}} \right)}^T},{{\left( {{\bf{\bar f}}_2^{{\rm{RF}}}} \right)}^T}, \cdots {{\left( {{\bf{\bar f}}_{{N_{{\rm{RF}}}}}^{{\rm{RF}}}} \right)}^T}} \right]^T}}$, and set the probability parameter ${\mathbf{u}={{\left[ {{u}_{1}},{{u}_{2}},\cdots ,{{u}_{N}} \right]}^{T}}}$ as an ${N \times 1}$ vector, where ${0\le {{u}_{n}}\le 1}$ presents the probability that ${{f_n} = 1/\sqrt N }$, ${{{f}_{n}}}$ is the ${n}$th element of ${\mathbf{f}}$.  Then, by initializing ${{{\mathbf{u}}^{\left( 0 \right)}}=\frac{1}{2}\times {{\mathbf{1}}_{N\times 1}}}$ (${{\bf{1}}}$ is the all-one vector), we assume that all the ${N}$ nonzero elements of ${{{\mathbf{F}}_{\text{RF}}}}$ belong to ${\frac{1}{\sqrt{N}}\left\{ -1,+1 \right\}}$ with equal probability, since no priori information is available. During the ${i}$th iteration, in step 1, we first generate ${S}$ candidate analog beamformers ${\left\{ \mathbf{F}_{\text{RF}}^{s} \right\}_{s=1}^{S}}$ based on the probability distribution ${\Xi \left( \mathcal{F};{{\mathbf{u}}^{\left( i \right)}} \right)}$ (i.e., generate ${\left\{ {{\mathbf{f}}^{s}} \right\}_{s=1}^{S}}$ according to ${{{\mathbf{u}}^{\left( i \right)}}}$, and reshape them as matrices belong to ${\mathcal{F}}$). Then, in step 2, we calculate the corresponding digital precoder ${\mathbf{F}_{\text{BB}}^{s}}$ according to the effective channel ${\mathbf{H}_{\text{eq}}^{s}=\mathbf{HF}_{\text{RF}}^{s}}$ for ${1\le s\le S}$. Note that there are lots of advanced digital precoder schemes~\cite{marzetta10}. In this paper, we adopt the classical ZF digital precoder with the near-optimal performance and low complexity as the example~\cite{marzetta10}, and ${\mathbf{F}_{\text{BB}}^{s}}$ can be computed as
\begin{equation}\label{eq12}
{{\bf{G}}^s} = {\left( {{\bf{H}}_{{\rm{eq}}}^s} \right)^H}{\left( {{\bf{H}}_{{\rm{eq}}}^s{{\left( {{\bf{H}}_{{\rm{eq}}}^s} \right)}^H}} \right)^{ - 1}}, \quad {\bf{F}}_{{\rm{BB}}}^s = {\beta ^s}{{\bf{G}}^s},
\end{equation}
where ${{\beta ^s} = \sqrt \rho /{\left\| {{\bf{F}}_{{\rm{RF}}}^s{{\bf{G}}^s}} \right\|_F}}$ is power normalized factor.

\begin{algorithm}[tp]
\caption{The proposed ACE-based hybrid precoding}
\KwIn{Channel matrix ${\mathbf{H}}$; Number of iterations ${I}$;
 \\\hspace*{+9.5mm} Number of candidates ${S}$; Number of elites ${{{S}_{\text{elite}}}}$.}
\textbf{Initialization}: ${i = 0}$; ${{{\mathbf{u}}^{\left( 0 \right)}}=\frac{1}{2}\times {{\mathbf{1}}_{N\times 1}}}$.
\\\textbf{for} ${0 \le i \le I}$
 \\1. Randomly generate ${S}$ candidate analog beamformers
 \\\hspace*{+3mm}${\left\{ \mathbf{F}_{\text{RF}}^{s} \right\}_{s=1}^{S}}$ based on ${\Xi \left( \mathcal{F};{{\mathbf{u}}^{\left( i \right)}} \right)}$;
 \\2. Compute ${S}$ corresponding digital precoders ${\left\{ \mathbf{F}_{\text{BB}}^{s} \right\}_{s=1}^{S}}$
 \\\hspace*{+2.5mm} based on the effective channel ${\mathbf{H}_{\text{eq}}^{s}=\mathbf{HF}_{\text{RF}}^{s}}$;
 \\3. Calculate the achievable sum-rate ${\left\{ {R\left( {{\bf{F}}_{{\rm{RF}}}^s} \right)} \right\}_{s = 1}^S}$~(\ref{eq9});
 \\4. Sort ${\left\{ {R\left( {{\bf{F}}_{{\rm{RF}}}^s} \right)} \right\}_{s = 1}^S}$ in a descend order as
 \\\hspace*{+11mm}${R\left( \mathbf{F}_{\text{RF}}^{\left[ 1 \right]} \right)\ge R\left( \mathbf{F}_{\text{RF}}^{\left[ 2 \right]} \right)\ge \cdots \ge R\left( \mathbf{F}_{\text{RF}}^{\left[ S \right]} \right)}$;
 \\5. Select elites as ${\mathbf{F}_{\text{RF}}^{\left[ 1 \right]},\mathbf{F}_{\text{RF}}^{\left[ 2 \right]},\cdots ,\mathbf{F}_{\text{RF}}^{\left[ {{S}_{\text{elite}}} \right]}}$;
 \\6. Calculate weight ${{{w}_{s}}}$ for each elite ${\mathbf{F}_{\text{RF}}^{\left[ s \right]}}$, ${1\le s\le {{S}_{\text{elite}}}}$;
 \\7. Update ${{{\mathbf{u}}^{\left( i+1 \right)}}}$ according to ${\left\{ {{w}_{s}} \right\}_{s=1}^{{{S}_{\text{elite}}}}}$ and ${\left\{ \mathbf{F}_{\text{RF}}^{\left[ s \right]} \right\}_{s=1}^{{{S}_{\text{elite}}}}}$;
 \\8. ${i = i + 1}$;
\\\textbf{end for}
\\\KwOut{Analog beamformer ${\mathbf{F}_{\text{RF}}^{\left[ 1 \right]}}$; Digital precoder ${\mathbf{F}_{\text{BB}}^{\left[ 1 \right]}}$.}
\end{algorithm}

After that, in step 3, the achievable sum-rate ${\left\{ {R\left( {{\bf{F}}_{{\rm{RF}}}^s} \right)} \right\}_{s = 1}^S}$ are calculated by substituting ${\mathbf{F}_{\text{RF}}^{s}}$  and ${\mathbf{F}_{\text{BB}}^{s}}$ (also a function of ${\mathbf{F}_{\text{RF}}^{s}}$) into~(\ref{eq9}). We sort ${\left\{ {R\left( {{\bf{F}}_{{\rm{RF}}}^s} \right)} \right\}_{s = 1}^S}$ in a descend order in step 4. Then, the elites can be obtained in step 5. In the conventional CE algorithm, the next step is to using elites to update ${{{\mathbf{u}}^{\left( i+1 \right)}}}$ by minimizing CE, which is equivalent to~\cite{rubinstein2013cross}
\begin{equation}\label{eq31}
{{\bf{u}}^{\left( {i + 1} \right)}} = \arg \mathop {\max }\limits_{{{\bf{u}}^{\left( i \right)}}} \frac{1}{S}\sum\limits_{s = 1}^{{S_{{\rm{elite}}}}} {\ln \Xi \left( {{\bf{F}}_{{\rm{RF}}}^{\left[ s \right]};{{\bf{u}}^{\left( i \right)}}} \right)} ,
\end{equation}
where ${\Xi \left( {{\bf{F}}_{{\rm{RF}}}^{\left[ s \right]};{{\bf{u}}^{\left( i \right)}}} \right)}$ denotes the probability to generate ${{{\bf{F}}_{{\rm{RF}}}^{\left[ s \right]}}}$. As mentioned above, in~(\ref{eq31}), the contributions of all elites are treated as the same, leading to performance degradation. To solve this problem, we propose to weight each elite adaptively based on its achievable sum-rate. Specifically, we first define an auxiliary parameter ${T}$ presenting the average achievable sum-rate of all elites as
\begin{equation}\label{eq13}
T=\frac{1}{{{S}_{\text{elite}}}}\sum\nolimits_{s=1}^{{{S}_{\text{elite}}}}{R\left( \mathbf{F}_{\text{RF}}^{\left[ s \right]} \right)}.
\end{equation}
Then, the weight ${{{w}_{s}}}$ of the elite ${\mathbf{F}_{\text{RF}}^{\left[ s \right]}}$ can be calculated in step 6 as ${{{w}_{s}}=R\left( \mathbf{F}_{\text{RF}}^{\left[ s \right]} \right)/T}$. Based on ${\left\{ {{w}_{s}} \right\}_{s=1}^{{{S}_{\text{elite}}}}}$,~(\ref{eq31}) can be modified as
\begin{equation}\label{eq14}
{{\bf{u}}^{\left( {i + 1} \right)}} = \arg \mathop {\max }\limits_{{{\bf{u}}^{\left( i \right)}}} {\mkern 1mu} \frac{1}{S}\sum\limits_{s = 1}^{{S_{{\rm{elite}}}}} {{w_s}\ln \Xi \left( {{\bf{F}}_{{\rm{RF}}}^{\left[ s \right]};{{\bf{u}}^{\left( i \right)}}} \right)}.
\end{equation}
Note that ${\Xi \left( \mathbf{F}_{\text{RF}}^{\left[ s \right]};{{\mathbf{u}}^{\left( i \right)}} \right)=\Xi \left( {{\mathbf{f}}^{\left[ s \right]}};{{\mathbf{u}}^{\left( i \right)}} \right)}$, and the ${n}$th element ${f_{n}^{\left[ s \right]}}$ of ${{{\mathbf{f}}^{\left[ s \right]}}}$ is a Bernoulli random variable, where ${f_n^{\left[ s \right]} = 1/\sqrt N }$ with probability ${u_{n}^{\left( i \right)}}$ and ${f_n^{\left[ s \right]} =  - 1/\sqrt N }$ with probability ${1-u_{n}^{\left( i \right)}}$. Therefore, we have
\begin{equation}\label{eq15}
\Xi \left( {{\bf{F}}_{{\rm{RF}}}^{\left[ s \right]};{{\bf{u}}^{\left( i \right)}}} \right)\! = \!\prod\limits_{n = 1}^N {{{\left( {u_n^{\left( i \right)}} \right)}^{\frac{1}{2}\left( {1\! +\! \sqrt N f_n^{\left[ s \right]}} \right)}}} {\left( {1\! -\! u_n^{\left( i \right)}} \right)^{\frac{1}{2}\left( {1\! -\! \sqrt N f_n^{\left[ s \right]}} \right)}}.
\end{equation}
After substituting~(\ref{eq15}) into~(\ref{eq14}), the first derivative of the target in~(\ref{eq14}) with respect to ${u_{n}^{\left( i \right)}}$ can be derived as
\begin{equation}\label{eq16}
\frac{1}{S}\sum\limits_{s = 1}^{{S_{{\rm{elite}}}}} {{w_s}\left( {\frac{{1 + \sqrt N f_n^{\left[ s \right]}}}{{2u_n^{\left( i \right)}}} - \frac{{1 - \sqrt N f_n^{\left[ s \right]}}}{{2\left( {1 - u_n^{\left( i \right)}} \right)}}} \right)} .
\end{equation}
Setting~(\ref{eq16}) to zero, ${{{\mathbf{u}}^{\left( i+1 \right)}}}$ can be updated in step 7 as
\begin{equation}\label{eq17}
u_n^{\left( {i + 1} \right)} = \frac{{\sum\nolimits_{s = 1}^{{S_{{\rm{elite}}}}} {{w_s}\left( {\sqrt N f_n^{\left[ s \right]} + 1} \right)} }}{{2\sum\nolimits_{s = 1}^{{S_{{\rm{elite}}}}} {{w_s}} }}.
\end{equation}
Such procedure above will be repeated (${i = i + 1}$ in step 8) until the maximum number of iterations ${I}$ is reached, and the analog beamformer and digital precoder will be selected as ${\mathbf{F}_{\text{RF}}^{\left[ 1 \right]}}$ and ${\mathbf{F}_{\text{BB}}^{\left[ 1 \right]}}$, respectively. Finally, it is worth pointing out that the proposed ACE-based hybrid precoding scheme can be also extended to the scenario where users employ multiple antennas. In this case, the analog beamformers at the BS and users should be jointly searched by the ACE algorithm.

\subsection{Computational complexity analysis}\label{S3.3}
In this subsection, the computational complexity of the proposed ACE-based hybrid precoding scheme is discussed.

From \textbf{Algorithm 1}, we can observe that the complexity of the ACE-based hybrid precoding scheme mainly comes from steps 2, 3, 6, and 7. In step 2, we need to compute ${S}$ effective channel matrices ${\left\{ {{\bf{H}}_{{\rm{eq}}}^s} \right\}_{s = 1}^S}$ and digital precoders ${\left\{ \mathbf{F}_{\text{BB}}^{s} \right\}_{s=1}^{S}}$ according to~(\ref{eq12}). Therefore, the complexity of this part is ${{\cal O}\left( {SN{K^2}} \right)}$. In step 3, the achievable sum-rate of each candidate is computed. Since we employ the digital ZF precoder, the SINR ${\gamma _k^s}$ of the ${k}$th user for the ${s}$th candidate  is simplified to ${\gamma _k^s = {\left( {{\beta ^s}/\sigma } \right)^2}}$. As a result, this part only involves the complexity ${{\cal O}\left( {S} \right)}$. In step 6, we calculate ${{{S_{{\rm{elite}}}}}}$ weights based on~(\ref{eq13}), which is quite simple with the complexity ${{\cal O}\left( {{S_{{\rm{elite}}}}} \right)}$. Finally, in step 7, the probability parameter ${{{\bf{u}}^{\left( {i + 1} \right)}}}$ is updated according to~(\ref{eq17}) with the complexity ${{\cal O}\left( {N{S_{{\rm{elite}}}}} \right)}$.

In summary, after ${I}$ iterations, the total computational complexity of the proposed ACE hybrid precoding scheme is ${{\cal O}\left( {ISN{K^2}} \right)}$. Since ${K}$ is usually small, ${I}$ and ${S}$ also do not have to be very large as will be verified in the next section,  we can conclude that the complexity of the proposed ACE-based hybrid precoding scheme is acceptable, which is comparable with the simple least squares (LS) algorithm.

\begin{figure}[tp]
\begin{center}
\vspace*{-4mm}\includegraphics[width=0.95\linewidth]{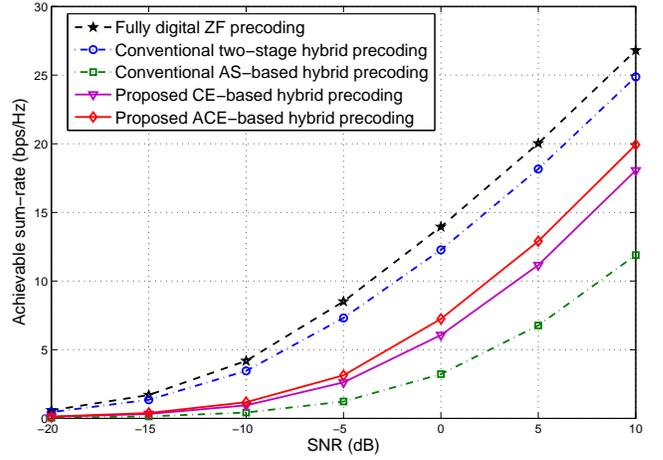}
\end{center}
\vspace*{-4mm}\caption{Achievable sum-rate comparison.} \label{FIG3}
\end{figure}

\begin{figure}[tp]
\begin{center}
\vspace*{0mm}\includegraphics[width=0.95\linewidth]{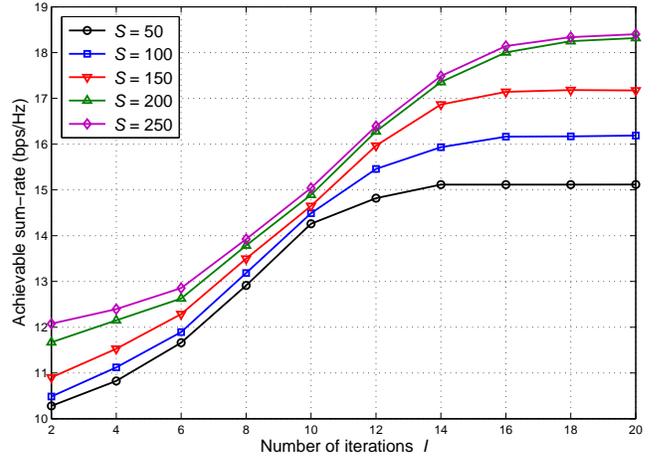}
\end{center}
\vspace*{-3mm}\caption{Achievable sum-rate against  ${S}$ and ${I}$.} \label{FIG3}
\end{figure}

\section{Simulation Results}\label{S4}
In this section, we provide the simulation results in terms of achievable sum-rate and energy-efficiency to evaluate the performance of the proposed ACE-based hybrid precoding scheme. The simulation parameters are described as follows: We assume that the BS employs an UPA with antenna spacing ${{d_1} = {d_2} = \lambda /2}$. For the ${k}$th user, we generate the channel ${{{\bf{h}}_k}}$ based on~(\ref{eq2}), where we assume: 1) ${{L_k} = 3}$; 2) ${\alpha _k^{\left( l \right)}\sim{\cal C}{\cal N}\left( {0,1} \right)}$ for ${1 \le l \le {L_k}}$; 3) ${\varphi _k^{\left( l \right)}}$ and ${\theta _k^{\left( l \right)}}$ follow the uniform distribution ${{\cal U}\left( { - \pi,\pi} \right)}$ for ${1 \le l \le {L_k}}$~\cite{sayeed2013beamspace}. Finally, the signal-to-noise ratio (SNR) is defined as ${\rho /{\sigma ^2}}$.

\begin{figure}[tp]
\begin{center}
\vspace*{-2mm}\includegraphics[width=0.95\linewidth]{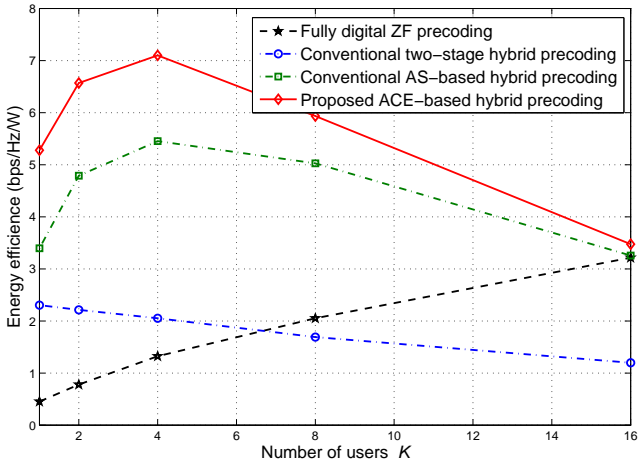}
\end{center}
\vspace*{-4mm}\caption{Energy efficiency comparison.} \label{FIG3}
\end{figure}

Fig. 2 shows the achievable sum-rate comparison in a typical mmWave massive MIMO system with ${N = 64}$, ${{N_{{\rm{RF}}}} = K = 4}$. In Fig. 2, the proposed CE-based (i.e., using the conventional CE algorithm to solve~(\ref{eq8})) and ACE-based hybrid precoding schemes are designed for SI-based architecture, where we set ${S = 200}$, ${{S_{{\rm{elite}}}}   = 40}$, and ${I = 20}$ for \textbf{Algorithm 1}, the conventional two-stage hybrid precoding scheme is designed for PS-based architecture with 4-bit phase shifters~\cite{alkhateeb2015limited}, and the conventional antenna selection (AS)-based hybrid precoding scheme is designed for SW-based architecture~\cite{mendez2015channel} with switches. Firstly, we can observe from Fig. 2 that the proposed ACE algorithm outperforms the traditional CE algorithm, where the SNR gap is about 1 dB. Note that the ACE algorithm only involves one additional step (i.e., step 6 in \textbf{Algorithm 1}) with negligible complexity. Therefore, the proposed ACE algorithm is more efficient. Moreover, Fig. 2 also shows that the proposed ACE-based hybrid precoding can achieve the sum-rate much higher than the conventional AS-based hybrid precoding, as it can achieve the potential array gains in mmWave massive MIMO systems. Finally, we can observe that the performance gap between ACE-based hybrid precoding and two-stage hybrid precoding is limited and keeps constant, which further verifies the conclusion  in \textbf{Lemma 1}.

Fig. 3 shows the achievable sum-rate of the proposed ACE-based hybrid precoding against the number of candidates ${S}$ and the number of iterations ${I}$, when ${{{S_{{\rm{elite}}}}/S = 0.2}}$, ${N = 64}$, ${{N_{{\rm{RF}}}} = K = 4}$, and SNR = 10 dB. From Fig. 3, we can observe that when ${S}$ is small, increasing ${S}$ will lead to an obvious improvement in the sum-rate performance. However, when ${S}$ is sufficiently large, such trend is no longer obvious. This indicates that the number of candidates ${S}$ does not have to be very large, e.g., ${S = 200}$ is enough. Furthermore, Fig. 3 also indicates that the proposed ACE-based hybrid precoding can converge with a small number of iterations, e.g., ${I = 20}$. These observations verify the rationality of the parameters for  \textbf{Algorithm 1} we used in Fig. 2.

Fig. 4 shows the energy efficiency comparison when ${N = 64}$ is fixed and ${{N_{{\rm{RF}}}} = K}$ varies from 1 to 16. The parameters for \textbf{Algorithm 1} are the same as Fig. 2. According to~\cite{gao15energy,mendez2016hybrid}, the energy efficiency can be defined as the ratio between the achievable sum-rate and the energy consumption, which should be~(\ref{eq4}),~(\ref{eq5}), and~(\ref{eq6}) for two-stage hybrid precoding, AS-based hybrid precoding, and ACE-based hybrid precoding, respectively. In this paper, we adopt the practical values ${\rho  = 30{\rm{mW}}}$~\cite{gao15energy}, ${{P_{{\rm{RF}}}} = 300{\rm{mW}}}$~\cite{gao15energy}, ${{P_{{\rm{BB}}}} = 200{\rm{mW}}}$~\cite{mendez2016hybrid}, ${{P_{{\rm{PS}}}} = 40{\rm{mW}}}$ (4-bit phase shifter)~\cite{mendez2016hybrid}, and ${{P_{{\rm{SW}}}} = {P_{{\rm{IN}}}} = 5{\rm{mW}}}$~\cite{mendez2016hybrid}. From Fig. 4, we can observe that the proposed ACE-based hybrid precoding with SI-based architecture can achieve much higher energy efficiency than the others, especially when ${K}$ is not very large (e.g., ${K \le 12}$). Furthermore, it is interesting to observe that when ${K \ge 8}$, the energy efficiency of the two-stage hybrid precoding with PS-based architecture is even lower than that of the fully digital ZF precoding. This is due to the fact that as ${K}$ grows, the number of phase shifters in PS-based architecture increases rapidly. As a result, the energy consumption of the phase shifter network will be huge, even higher than that of RF chains.

\section{Conclusions}\label{S5}
In this paper, we propose an energy-efficient SI-based hybrid precoding architecture, where the analog part is realized by a small number of switches and inverters. The performance analysis proves that the performance gap between the proposed SI-based architecture and the traditional near-optimal one keeps constant and limited. Then, by employing the idea of CE optimization in machine learning, we further propose an ACE-based hybrid precoding scheme with low complexity for SI-based architecture. Simulation results verify that our scheme can achieve the satisfying sum-rate performance and much higher energy efficiency than traditional schemes.


\bibliography{IEEEabrv,Gao1Ref}

\end{document}